\documentclass[preprint,amsfonts,amssymb,amsmath,eqsecnum,%
tightenlines,nofootinbib,showpacs,showkeys]{revtex4}

\newcommand{\cN}{\mathcal{N}}
\newcommand{\cV}{\mathcal{V}}
\newcommand{\cW}{\mathcal{W}}
\newcommand{\bH}{\mathbf{H}}
\newcommand{\bQ}{\mathbf{Q}}
\newcommand{\tH}{\tilde{H}}
\newcommand{\tP}{\tilde{P}}
\newcommand{\tcV}{\tilde{\cV}}
\newcommand{\tw}{\tilde{w}}
\newcommand{\tvph}{\tilde{\varphi}}

\newcommand{\bbR}{\mathbb{R}}
\newcommand{\bbC}{\mathbb{C}}
\newcommand{\rmi}{\mathrm{i}}
\newcommand{\rme}{\mathrm{e}}
\newcommand{\rmd}{\mathrm{d}}
\newcommand{\del}{\partial}
\newcommand{\rnu}{\sqrt{\nu}}
\newcommand{\braket}[1]{\bigl\langle{#1}\bigr\rangle}

\begin{document}


\title{$\cN$-fold Supersymmetry in Quantum Systems with
 Position-dependent Mass}
\author{Toshiaki Tanaka}
\email{ttanaka@mail.tku.edu.tw}
\affiliation{Department of Physics, Tamkang University,\\
 Tamsui 25137, Taiwan, R.O.C.}


\begin{abstract}

We formulate the framework of $\cN$-fold supersymmetry in
one-body quantum mechanical systems with position-dependent
mass (PDM). We show that some of the significant properties
in the constant-mass case such as the equivalence to weak
quasi-solvability also hold in the PDM case. We develop a
systematic algorithm for constructing an $\cN$-fold
supersymmetric PDM system. We apply it to obtain type A
$\cN$-fold supersymmetry in the case of PDM, which is
characterized by the so-called type A monomial space.
The complete classification and general form of effective
potentials for type A $\cN$-fold supersymmetry in the PDM case
are given.

\end{abstract}






\maketitle

\section{Introduction}
\label{sec:intro}

Recently, quantum systems with position-dependent mass (PDM) have
attracted much attention in various research fields of physics
such as semiconductors, quantum dots, liquid crystals, and so on.
Accordingly, investigations into exact solutions of PDM quantum
systems have been carried out increasingly in the last few years.
For references, see e.g., Refs.~\cite{DCH98,DCH99,MI99,PRCGP99,DA00,%
Al02,GGTO02,GOGU02,KKK02,RR02,DHA03,KK03,KT03,SL03,YDS04,YD04,Ch04,%
CC04,QT04,BGQR04a,GK04,BBQT05,DL05,RR05,BGQR05,Qu05} and those
cited therein.
Due to the fact that the position-dependent mass $m(q)$ does not commute
with the momentum operator $p=-\rmi\rmd/\rmd q$, ambiguity arises in
defining a quantum kinetic operator which is formally Hermitian and
reduces to the classical kinetic term $T=p^{2}/2m(q)$. Hence,
the following operator proposed by von Roos \cite{vR83} has been
generally considered:
\begin{align}
H=-\frac{1}{4}\left(m(q)^{\alpha}\frac{\rmd}{\rmd q}m(q)^{\beta}
 \frac{\rmd}{\rmd q}m(q)^{\gamma}+m(q)^{\gamma}\frac{\rmd}{\rmd q}
 m(q)^{\beta}\frac{\rmd}{\rmd q}m(q)^{\alpha}\right)+V(q),
\end{align}
with the constraint $\alpha+\beta+\gamma=-1$. A different choice
of the parameters results in a different correction to the original
potential profile $V(q)$, and the above Hamiltonian always has
the following form:
\begin{align}
H=-\frac{1}{2m(q)}\frac{\rmd^{2}}{\rmd q^{2}}
 +\frac{m'(q)}{2m(q)^{2}}\frac{\rmd}{\rmd q}+U(q),
\label{eq:gPDMH}
\end{align}
where the \emph{effective} potential $U(q)$ is given by
\begin{align}
U(q)=V(q)-(\alpha+\gamma)\frac{m''(q)}{4m(q)^{2}}
 +(\alpha\gamma+\alpha+\gamma)\frac{m'(q)^{2}}{2m(q)^{3}}.
\end{align}
Hence, the typical investigations into (quasi-)exact solvability of
PDM quantum systems consist in finding simultaneously a pair of an
effective potential $U(q)$ and a mass function $m(q)$ for which the
PDM Hamiltonian \eqref{eq:gPDMH} admits (a number of) exact
eigenfunctions in closed form. Up to now, two different methods have
been frequently employed, namely, coordinate transformations
including point canonical transformations \cite{DCH98,DCH99,DA00,%
Al02,GGTO02,GOGU02,KKK02,KK03,YD04,CC04,BGQR04a,BGQR05}, and
supersymmetric methods \cite{MI99,PRCGP99,GGTO02,GOGU02,DHA03,KT03,%
SL03,QT04,BGQR04a,GK04,BBQT05,RR05,BGQR05}. The latter approaches
were also applied to many-body PDM quantum systems \cite{Qu05}. For
the methods themselves developed in ordinary quantum systems, see
references cited therein.

On the other hand, the framework of $\cN$-fold supersymmetry has
been rapidly developed in one-body ordinary quantum mechanical
systems, especially in the last few years.
It was originally proposed as a higher derivative generalization
of ordinary supercharges \cite{AIS93}. Later, a significant
breakthrough was achieved by the proof of the equivalence between
$\cN$-fold supersymmetry and (weak) quasi-solvability \cite{AST01b}.
Based on the equivalence, a systematic algorithm for constructing
an $\cN$-fold supersymmetric system was developed \cite{GT05}.
Owing to the fact that the framework of $\cN$-fold supersymmetry
includes all the ordinary supersymmetric methods as its special cases
and to its equivalence to weak quasi-solvability, which is less
restrictive concept than quasi-exact solvability \cite{TU87,Us94},
it provides one of the most powerful methods for investigating
one-body quantum mechanical systems which admit analytic solutions
in so far the least restrictive sense.

Considering the present situation described above, it is natural
to ask whether or not we can formulate $\cN$-fold supersymmetry
also in PDM quantum systems, especially as a powerful tool for
constructing (quasi-)solvable PDM systems. In this paper, we show
that it is indeed possible and that the characteristic properties
such as the equivalence to weak quasi-solvability also hold in the
PDM case. Furthermore, we generalize the systematic algorithm for
constructing an $\cN$-fold supersymmetric ordinary quantum system in
Ref.~\cite{GT05} to the PDM case. As an illustration, we apply it
to construct so called type A $\cN$-fold supersymmetry
\cite{AST01a,Ta03a} in PDM quantum systems.

The paper is organized as follows. In the following section, we review
the precise definition of some important concepts in the paper,
such as quasi-solvability, to avoid ambiguity. In
Section~\ref{sec:nsusy}, we define $\cN$-fold supersymmetry in
one-body quantum systems with position-dependent mass and discuss
a couple of general significant consequences. In
Section~\ref{sec:algo}, we develop a systematic algorithm for
constructing an $\cN$-fold supersymmetric PDM quantum system by
slightly generalizing the one in the constant-mass case in
Ref.~\cite{GT05}. In Section~\ref{sec:typea}, we apply the algorithm
to construct type A $\cN$-fold supersymmetry in the PDM case.
In Section~\ref{sec:class}, we completely classify and present
the explicit forms of all the inequivalent type A models with
arbitrary position dependence of mass. In the last section, we
summarize the results and discuss various future issues.

\section{Definition}
\label{sec:def}

First of all, we shall give precise definition of some key concepts
in the paper. The original ideas can be found in Refs.~\cite{AST01b,%
Tu94} but we have slightly modified the terms from the view point
of the recent advances in this research field.
Let $H$ be a linear differential operator of a single
variable  and $P_{\cN}$ be an (at most) $\cN$th-order linear
differential operator. Then, $H$ is said to be \emph{weakly
quasi-solvable} with respect to $P_{\cN}$ \cite{Ta03a} if it preserves
the vector space $\cV_{\cN}$ defined by
\begin{align}
\cV_{\cN}=\ker P_{\cN}.
\label{eq:cvker}
\end{align}
A linear differential operator $H$ of a single variable is said to
be \emph{quasi-solvable} if it preserves a finite-dimensional
functional space $\cV_{\cN}$ whose basis admits an analytic expression
in closed form:
\begin{align}
H\cV_{\cN}&\subset\cV_{\cN}, & \dim\cV_{\cN}&=n(\cN)<\infty,
 & \cV_{\cN}&=\braket{\phi_{1}(q),\dots,\phi_{n(\cN)}(q)}.
\label{eq:defqs}
\end{align}
It is evident that a weakly quasi-solvable operator is also
quasi-solvable if linearly independent $\cN$ solutions of
$P_{\cN}\phi=0$ can be obtained in closed form. An immediate
consequence of the above definition of quasi-solvability is that,
since we can calculate finite-dimensional matrix elements
$\mathsf{H}_{k,l}$ defined by
\begin{align}
H\phi_{k}=\sum_{l=1}^{n(\cN)}\mathsf{H}_{k,l}\phi_{l},\qquad
 k=1,\dots,n(\cN),
\label{eq:defbS}
\end{align}
we can diagonalize the operator $H$ and obtain its spectra
in the space $\cV_{\cN}$ with finite algebraic manipulations. However,
these calculable spectra and the corresponding vectors of $\cV_{\cN}$
in general only give \emph{local} solutions of the characteristic
equation. This fact naturally leads to the well-known concept of
quasi-exact solvability. A quasi-solvable operator $H$ is said to be
\emph{quasi-exactly solvable} (on $S\subset\bbR$ or $\bbC$) if
the invariant space $\cV_{\cN}$ is a subspace of a Hilbert space
$L^{2}(S)$ on which the operator $H$ is naturally defined.
It is evident that if an operator is not only quasi-solvable but
also quasi-exactly solvable,  the calculable spectra and the
corresponding vectors of $\cV_{\cN}$ give a part of the \emph{exact}
eigenvalues and eigenfunctions of $H$, respectively.
On the other hand, there are cases in which all the eigenfunctions
and eigenvalues can be obtained analytically such as the well-known
harmonic oscillator. In order to characterize these cases, we
first introduce another subclass of quasi-solvability.
A quasi-solvable operator $H$ is said to be \emph{solvable}
if it preserves an infinite flag of finite-dimensional functional
spaces $\cV_{\cN}$,
\begin{align}
\cV_{1}\subset\cV_{2}\subset\dots\subset\cV_{\cN}\subset\cdots,
\label{eq:flagv}
\end{align}
whose bases admit explicit analytic expressions in closed form,
that is,
\begin{align}
H\cV_{\cN}&\subset\cV_{\cN}, & \dim\cV_{\cN}&=n(\cN)<\infty,
 & \cV_{\cN}&=\braket{\phi_{1}(q),\dots,\phi_{n(\cN)}(q)},
\label{eq:defsv}
\end{align}
for all $\cN=1,2,3,\ldots$. A consequence of solvability defined
above is that, for an arbitrary natural number $\cN$, we can
obtain additional $n(\cN+1)-n(\cN)$ local solutions of
the characteristic equation for $H$ in $\cV_{\cN+1}$ with
finite algebraic manipulations, based on the knowledge of
them in the subspace $\cV_{\cN}\subset\cV_{\cN+1}$.
Hence, arbitrary number of local solutions of the characteristic
equation for $H$ can be calculated, in principle, with a well-defined
finite algebraic algorithm. As in the case of quasi-solvability,
however, they are not necessarily the exact eigenfunctions and
eigenvalues of $H$. A solvable operator $H$ is said to be
\emph{exactly solvable} (on $S\subset\bbR$ or $\bbC$) if the sequence
of the spaces \eqref{eq:flagv} in $S$ satisfies,
\begin{align}
\overline{\cV_{\cN}(S)}\rightarrow L^{2}(S)\qquad
 (\cN\rightarrow\infty).
\label{eq:limvs}
\end{align}
This definition precisely fits what is commonly meant by exactly
solvable. In the PDM case, it is pointed out \cite{BBQT05} that in
addition to the square integrability, any eigenfunction $\psi$ of
a PDM Hamiltonian \eqref{eq:gPDMH} should satisfy on the boundary
$\del S$
\begin{align}
\frac{|\psi(q)|^{2}}{m(q)^{\frac{1}{2}}}\rightarrow 0
 \qquad (q\rightarrow \del S),
\end{align}
in order for the PDM Hamiltonian to be Hermitian. In this paper,
however, we do not discuss Hilbert spaces, boundary conditions,
normalizability, etc. which strongly rely on a specific choice of
$m(q)$. We do not assume Hermiticity of Hamiltonians and thus reality
of effective potential $U(q)$ and mass function $m(q)$ either, taking
into account possible application to recently developing non-Hermitian
quantum theories.

\section{$\cN$-fold Supersymmetry in {PDM} Quantum Systems}
\label{sec:nsusy}

To define $\cN$-fold supersymmetry in one-body PDM quantum systems,
let us first introduce fermionic variables $\psi$ and
$\psi^{\dagger}$ satisfying
\begin{align}
\{\psi,\psi\}=\{\psi^{\dagger},\psi^{\dagger}\}=0,\qquad
 \{\psi,\psi^{\dagger}\}=1.
\end{align}
In this paper, we consider a super-Hamiltonian
\begin{align}
\bH_{\cN}=H_{\cN}^{-}\psi\psi^{\dagger}
 +H_{\cN}^{+}\psi^{\dagger}\psi,
\label{eq:defsH}
\end{align}
where the components $H_{\cN}^{\pm}$ are given by PDM
Schr\"odinger operators:
\begin{align}
H_{\cN}^{\pm}=-\frac{1}{2m(q)}\frac{\rmd^{2}}{\rmd q^{2}}
 +\frac{m'(q)}{2m(q)^{2}}\frac{\rmd}{\rmd q}+U_{\cN}^{\pm}(q).
\label{eq:PDMSo}
\end{align}
For the above system, we define $\cN$-fold supercharges
$\bQ_{\cN}^{\pm}$ by
\begin{align}
\bQ_{\cN}^{-}=P_{\cN}\psi^{\dagger}\equiv P_{\cN}^{-}
 \psi^{\dagger},\quad \bQ_{\cN}^{+}=P_{\cN}^{t}\psi\equiv
 P_{\cN}^{+}\psi,
\label{eq:defQN}
\end{align}
where the operator $P_{\cN}$ is an $\cN$th-order linear differential
operator of the following form:\footnote{Here we fix the irrelevant
overall multiplicative constant factor so that $P_{\cN}$ is monic
when $m(q)=1$.}
\begin{align}
P_{\cN}=m(q)^{-\frac{\cN}{2}}\frac{\rmd^{\cN}}{\rmd q^{\cN}}
 +\sum_{k=0}^{\cN-1}w_{k}(q)\frac{\rmd^{k}}{\rmd q^{k}}.
\label{eq:defPN}
\end{align}
In Eq.~\eqref{eq:defQN}, the superscript $t$ denotes the formal
transposition. A system $\bH_{\cN}$ is said to be \emph{$\cN$-fold
supersymmetric} with respect to $\bQ_{\cN}^{\pm}$ if it commutes
with them:
\begin{align}
\bigl[\bQ_{\cN}^{\pm},\bH_{\cN}\bigr]=0.
\end{align}
From the above simple generalization of $\cN$-fold supersymmetry to
PDM systems, we can show that most of the relevant consequences of
$\cN$-fold supersymmetry in ordinary quantum systems with constant
mass also hold in the present case. Let us first prove the equivalence
between $\cN$-fold supersymmetry and weak quasi-solvability. Suppose
a super-Hamiltonian $\bH_{\cN}$ is $\cN$-fold supersymmetric.
It is easy to see that the component operators $H_{\cN}^{\pm}$
satisfy the following intertwining relation
\begin{align}
P_{\cN}^{-}H_{\cN}^{-}=H_{\cN}^{+}P_{\cN}^{-},
\label{eq:inter}
\end{align}
and its formal transposition, $P_{\cN}^{+}H_{\cN}^{+}=H_{\cN}^{-}
P_{\cN}^{+}$,
and thus they preserve the vector spaces $\ker P_{\cN}^{\pm}$,
respectively. Hence, $H_{\cN}^{\pm}$ are weakly quasi-solvable.
Conversely, let us assume that a PDM Schr\"odinger operator $H$
is weakly quasi-solvable with respect to an $\cN$th-order linear
differential operator $P_{\cN}$ of the form \eqref{eq:defPN}.
Define a linear operator $G=P_{\cN}H-H_{1}P_{\cN}$ with
\begin{align}
H_{1}=H+m^{\frac{\cN-2}{2}}w'_{\cN-1}+\frac{\cN-1}{2}
 m^{\frac{\cN-4}{2}}m'w_{\cN-1}+\frac{\cN^{\,2}m''}{4m^{2}}
 -\frac{3\cN^{\,2}(m')^{2}}{8m^{3}}.
\label{eq:diffH}
\end{align}
From the assumed weak quasi-solvability, $H\ker P_{\cN}\subset
\ker P_{\cN}$, we have
\begin{align}
\phi\in\ker P_{\cN}\quad\Longrightarrow\quad G\phi=P_{\cN}H\phi
 -H_{1}P_{\cN}\phi=0.
\label{eq:nullG}
\end{align}
On the other hand, it is easy to check that $G$ is of order (at
most) $\cN-1$ while the dimension of $\ker P_{\cN}$ is $\cN$.
Hence, Eq.~\eqref{eq:nullG} cannot be hold unless $G$ is a null
operator. Therefore, if we put $H_{\cN}^{-}=H$, $H_{\cN}^{+}=H_{1}$,
$P_{\cN}^{-}=P_{\cN}$, and $P_{\cN}^{+}=P_{\cN}^{t}$, they
satisfy the intertwining relation \eqref{eq:inter} and thus
compose an $\cN$-fold supersymmetric system.

Another significant consequence of $\cN$-fold supersymmetry in
ordinary quantum systems is that the anti-commutator of $\cN$-fold
supercharges $\bQ_{\cN}^{-}$ and $\bQ_{\cN}^{+}$ is a polynomial of
degree $\cN$ in the super-Hamiltonian $\bH_{\cN}$, the polynomial
being (proportional to) the characteristic polynomial of the component
Hamiltonians $H_{\cN}^{\pm}$ restricted in the invariant subspaces
$\cV_{\cN}^{\pm}$ \cite{AST01b,AS03}:
\begin{align}
\{\bQ_{\cN}^{-},\bQ_{\cN}^{+}\}\propto\det\left(H_{\cN}^{\pm}
 \bigr|_{\cV_{\cN}^{\pm}}-\bH_{\cN}\right).
\label{eq:anti}
\end{align}
We can easily prove that the same is also true in the PDM case.
Indeed, in the proof of the theorem on SUSY algebras with T symmetry
in Ref.~\cite{AS03}, we only need to modify the functions
$w_{\cN}^{\pm}$ in Eq.~(6) as $w_{\cN}^{\pm}(x)=(\mp 1)^{\cN}
m(x)^{-\frac{\cN}{2}}$ and the definition of the operators
$r_{l}^{\pm}$ in Eq.~(19) as
\begin{align}
r_{l}^{-}=m(x)^{-\frac{1}{2}}\biggl(\frac{\rmd}{\rmd x}+\chi_{l}^{-}
 (x)\biggr),\quad r_{l}^{+}=\biggl(-\frac{\rmd}{\rmd x}+\chi_{l}^{+}
 (x)\biggr)m(x)^{-\frac{1}{2}},\quad l=1,\dots,\cN,
\end{align}
for a proof of Eq.~\eqref{eq:anti} in the PDM case. In our definition
of Hamiltonians \eqref{eq:PDMSo}, the proportional constant in
Eq.~\eqref{eq:anti} is $2^{\cN}$.

Other characteristic features such as the almost isospectral property
between a pair of Hamiltonians $H_{\cN}^{\pm}$ are easily derived
in the same way as in the ordinary case (cf. Ref.~\cite{AST01b}), and
thus we don't repeat a presentation of them in this paper.

\section{Algorithm for constructing an $\cN$-fold {SUSY PDM} system}
\label{sec:algo}

In Ref.~\cite{GT05}, a systematic algorithm for constructing an
$\cN$-fold supersymmetric ordinary quantum mechanical system was
developed. In this section, we show that it can be easily generalized
to the PDM case. The starting point is an $\cN$-dimensional linear
functional space
\begin{align}
\tcV_{\cN}^{-}=\braket{\tvph_{1}(z),\dots,\tvph_{\cN}(z)},
\end{align}
and a second-order linear differential operator
\begin{align}
\tH_{\cN}^{-}=-A(z)\frac{\rmd^{2}}{\rmd z^{2}}-B(z)
 \frac{\rmd}{\rmd z}-C(z),
\label{eq:tH-}
\end{align}
which leaves $\tcV_{\cN}$ invariant. Let
\begin{align}
\tP_{\cN}^{-}=g(z)\biggl(\frac{\rmd^{\cN}}{\rmd z^{\cN}}
 +\sum_{k=0}^{\cN-1}\tw_{k}(z)\frac{\rmd^{k}}{\rmd z^{k}}\biggr)
\label{eq:tPN-}
\end{align}
denote the most general $\cN$th-order linear differential operator
with kernel $\tcV_{\cN}$, where the function $g(z)$ is for the time
being undetermined. We shall first construct another second-order
linear differential operator of the form
\begin{align}
\tH_{\cN}^{+}=\tH_{\cN}^{-}-\delta C(z),
\label{eq:defdC}
\end{align}
satisfying the intertwining relation
\begin{align}
\tP_{\cN}^{-}\tH_{\cN}^{-}-\tH_{\cN}^{+}\tP_{\cN}^{-}=0.
\label{eq:ginter}
\end{align}
To this end, note that the l.h.s. of Eq.~\eqref{eq:ginter}
is in general a linear differential operator of order $\cN+1$.
Equating to zero the coefficients of $\del_{z}^{\,\cN+1}$ and
$\del_{z}^{\,\cN}$ in this operator, we obtain the following two
equations for the functions $g(z)$ and $\delta C(z)$:
\begin{gather}
\frac{g'}{g}=\frac{\cN}{2}\frac{A'}{A},
\label{eq:cond1}\\
\delta C=\frac{\cN(\cN-2)}{2}\left(A''-\frac{(A')^{2}}{2A}\right)
 +\cN\left(B'-\frac{BA'}{2A}\right)-A'\tw_{\cN-1}-2A\tw'_{\cN-1}.
\label{eq:cond2}
\end{gather}
When Eqs.~\eqref{eq:cond1} and \eqref{eq:cond2} are satisfied,
the l.h.s. of Eq.~\eqref{eq:ginter} is a linear differential operator
of order at most $\cN-1$ annihilating the $\cN$-dimensional vector
space $\tcV_{\cN}$, and hence it vanishes identically.

The last step in our construction consists in applying a change of
variable
\begin{align}
z=z(q)
\end{align}
and a \emph{gauge} transformation
\begin{align}
\tH_{\cN}^{\pm}\mapsto\rme^{-\cW_{\cN}^{-}}\tH_{\cN}^{\pm}\,
 \rme^{\cW_{\cN}^{-}}\bigr|_{z=z(q)}\equiv H_{\cN}^{\pm},
\label{eq:Ham+-}
\end{align}
to simultaneously convert $\tH_{\cN}^{\pm}$ into the PDM Schr\"odinger
form \eqref{eq:PDMSo}. Note that it is certainly possible since (by
construction) $\tH_{\cN}^{-}$ and $\tH_{\cN}^{+}$ differ by a scalar
function only. The appropriate change of variable and gauge
transformation are determined by
\begin{gather}
z'(q)^{2}=2m(q)A(z),
\label{eq:cond3}\\
\frac{\rmd\cW_{\cN}^{-}}{\rmd q}=\frac{z''(q)}{2z'(q)}
 -\frac{m(q)B(z)}{z'(q)}-\frac{m'(q)}{2m(q)}.
\label{eq:cond4}
\end{gather}
The effective potentials $U_{\cN}^{\pm}$ are given by
\begin{align}
U_{\cN}^{\pm}(q)=\frac{1}{2m(q)}\left[\left(\frac{\rmd\cW_{\cN}^{-}}{
 \rmd q}\right)^{2}-\frac{\rmd^{2}\cW_{\cN}^{-}}{\rmd q^{2}}
 +\frac{m'(q)}{m(q)}\frac{\rmd\cW_{\cN}^{-}}{\rmd q}\right]
 -C^{\pm}(z(q)),
\end{align}
where $C^{-}(z)=C(z)$ and $C^{+}(z)=C(z)+\delta C(z)$. From the above
construction it immediately follows that the system \eqref{eq:defsH}
and \eqref{eq:defQN}, with $H_{\cN}^{\pm}$ given by Eq.~\eqref{eq:Ham+-}
and $P_{\cN}$ by
\begin{align}
P_{\cN}=\rme^{-\cW_{\cN}^{-}}\tP_{\cN}^{-}\,\rme^{\cW_{\cN}^{-}}
 \bigr|_{z=z(q)},
\label{eq:drvPN}
\end{align}
is $\cN$-fold supersymmetric. Indeed, intertwining relation
\eqref{eq:inter} follows by applying the gauge transformation and
change of variable to Eq.~\eqref{eq:ginter}. Furthermore, it is
important to note that the form of $P_{\cN}$ in Eq.~\eqref{eq:drvPN}
is compatible with Eq.~\eqref{eq:defPN}. From Eqs.~\eqref{eq:cond1}
and \eqref{eq:cond3} we have
\begin{align}
\frac{g'(z)}{g(z)}=\frac{\cN z''(q)}{z'(q)^{2}}
 -\frac{\cN m'(q)}{2m(q)z'(q)}.
\end{align}
Integrating the latter equation we obtain
\begin{align}
g(z)=m(q)^{-\frac{\cN}{2}}z'(q)^{\cN},
\label{eq:gofz}
\end{align}
where we take the proportional constant to be $1$. Thus, it
immediately follows from Eqs.~\eqref{eq:tPN-} and \eqref{eq:drvPN}
that $P_{\cN}$ in Eq.~\eqref{eq:drvPN} is indeed of the form
\eqref{eq:defPN}. It is evident from the construction that the
Hamiltonian $H_{\cN}^{-}$ preserves the kernel of $P_{\cN}$ given by
\begin{align}
\cV_{\cN}^{-}\equiv\ker P_{\cN}=\braket{\varphi_{1}(q),\dots,
 \varphi_{\cN}(q)},
\end{align}
with
\begin{align}
\varphi_{i}(q)=\rme^{-\cW_{\cN}^{-}}\tvph_{i}(z)\bigr|_{z=z(q)}
 \quad (i=1,\dots,\cN).
\end{align}
We thus call the space $\cV_{\cN}^{-}$ the \emph{solvable sector}
of $H_{\cN}^{-}$.

Although the construction of an $\cN$-fold supersymmetric PDM system
itself has been completed, we can make it entirely symmetric with
respect to the partner Hamiltonians $H_{\cN}^{-}$ and $H_{\cN}^{+}$.
For this purpose, note that from the transposition of the intertwining
relation \eqref{eq:inter}, it follows that $H_{\cN}^{+}$ leaves
invariant the kernel of the supercharge
\begin{align}
P_{\cN}^{+}=P_{\cN}^{t}=\rme^{\cW_{\cN}^{-}}\tP_{\cN}^{t}
 \,\rme^{-\cW_{\cN}^{-}}.
\end{align}
Using the identity $(\del_{z})^{t}=-z'(q)\del_{z}z'(q)^{-1}$ and
Eq.~\eqref{eq:tPN-} with \eqref{eq:gofz}, we can express $P_{\cN}^{+}$
as
\begin{align}
P_{\cN}^{+}=\rme^{-\cW_{\cN}^{+}}\bar{P}_{\cN}^{+}\,\rme^{\cW_{\cN}^{+}},
\end{align}
where
\begin{align}
\bar{P}_{\cN}^{+}=m(q)^{-\frac{\cN}{2}}z'(q)^{\cN}\left[\left(
 -\frac{\rmd}{\rmd z}\right)^{\cN}+\sum_{k=0}^{\cN-1}\left(
 -\frac{\rmd}{\rmd z}\right)^{k}\tw_{k}(z)\right],
\label{eq:bPN+}
\end{align}
and the function $\cW_{\cN}^{+}$ is given by
\begin{align}
\cW_{\cN}^{+}=-\cW_{\cN}^{-}+(\cN-1)\ln |z'(q)|-\frac{\cN}{2}\ln
 |m(q)|.
\label{eq:WN+}
\end{align}
Now, the partner Hamiltonians $H_{\cN}^{\pm}$ are expressed in
a completely symmetric way as
\begin{align}
H_{\cN}^{\pm}=\rme^{-\cW_{\cN}^{\pm}}\bar{\tH}_{\cN}^{\pm}\,
 \rme^{\cW_{\cN}^{\pm}},
\label{eq:sympH}
\end{align}
where the gauged Hamiltonians $\tH_{\cN}^{-}$ and $\bar{H}_{\cN}^{+}$
leave the kernels of the gauged $\cN$-fold supercharges $\tcV_{\cN}^{-}
=\ker\tP_{\cN}^{-}$ and $\bar{\cV}_{\cN}^{+}=\ker\bar{P}_{\cN}^{+}$,
respectively. To express $\cW_{\cN}^{\pm}$ in a symmetric way, we
introduce two functions by
\begin{align}
W(q)&=\frac{1}{2}\left(\frac{\rmd\cW_{\cN}^{-}}{\rmd q}-
 \frac{\rmd\cW_{\cN}^{+}}{\rmd q}\right),
\label{eq:defWq}\\
E(q)&=\frac{z''(q)}{z'(q)}.
\label{eq:defEq}
\end{align}
From Eq.~\eqref{eq:cond3}, its immediate consequence
\begin{align}
z''(q)=m(q)A'(z)+\frac{m'(q)A(z)}{z'(q)},
\end{align}
and Eq.~\eqref{eq:cond4}, the function $W(q)$ is expressed as
\begin{align}
W(q)=-\frac{m(q)}{z'(q)}\left(\frac{\cN-2}{2}A'(z)+B(z)\right)
 \equiv -\frac{m(q)}{z'(q)}Q(z).
\label{eq:defQz}
\end{align}
We then have
\begin{align}
\cW_{\cN}^{\pm}&=-\frac{\cN}{4}\ln |m(q)|+\frac{\cN-1}{2}\int\rmd q\,
 E(q)\mp\int\rmd q\, W(q)\notag\\
&=-\frac{1}{4}\ln|m(q)|+\frac{\cN-1}{4}\ln |2A(z)|\pm\int\rmd z\,
 \frac{m(q)Q(z)}{2A(z)}.
\label{eq:cWN+-}
\end{align}
The connection between the gauged Hamiltonians $\bar{H}_{\cN}^{+}$
and $\tH_{\cN}^{+}$ follows easily from Eqs.~\eqref{eq:Ham+-},
\eqref{eq:sympH}, and \eqref{eq:defWq} as
\begin{align}
\bar{H}_{\cN}^{+}=\exp\left(-2\int\rmd q\, W(q)\right)\tH_{\cN}^{+}
 \exp\left(2\int\rmd q\, W(q)\right).
\end{align}
Using Eqs.~\eqref{eq:tH-}, \eqref{eq:defdC}, \eqref{eq:cond2},
\eqref{eq:cond3}, and \eqref{eq:defQz}, we obtain the following
unified formula for the gauged Hamiltonians:
\begin{align}
\bar{\tH}_{\cN}^{\pm}=&\,-A(z)\frac{\rmd^{2}}{\rmd z^{2}}+\biggl[
 \frac{\cN-2}{2}A'(z)\pm Q(z)\biggr]\frac{\rmd}{\rmd z}-C(z)\notag\\
 &\,-(1\pm 1)\biggl[\frac{\cN-1}{2}Q'(z)-\frac{1}{2}A'(z)\tw_{\cN-1}
 (z)-A(z)\tw'_{\cN-1}(z)\biggr].
\label{eq:gsHam}
\end{align}
Interestingly, the form of the gauged Hamiltonian $\bar{H}_{\cN}^{+}$
as well as $\tH_{\cN}^{-}$ is completely the same as in the case
of constant mass, cf. Eq.~(2.45) in Ref.~\cite{GT05}.
It should also be noted that the mass dependence of the Hamiltonians
$H_{\cN}^{\pm}$ given by Eq.~\eqref{eq:sympH} emerges only through
the change of variable determined by Eq.~\eqref{eq:cond3} and
the gauge potentials by Eqs.~\eqref{eq:cond4} and \eqref{eq:WN+}
since the gauged Hamiltonians \eqref{eq:gsHam} do not depend on
the mass function $m(q)$. As a consequence, each of the spectrum of
the Hamiltonians $H_{\cN}^{\pm}$ does not depend on the mass function
$m(q)$.\footnote{Some of the eigenvalues can (dis)appear depending
on $m(q)$ since normalizability of the corresponding wave functions
does rely on it.} This kind of spectral independence from the
mass function is first pointed out in Ref.~\cite{RR02}.

\section{Type {A} $\cN$-fold Supersymmetry}
\label{sec:typea}

We shall now construct type A $\cN$-fold supersymmetric PDM quantum
systems with the aid of the algorithm just developed in the previous
section. Type A $\cN$-fold supersymmetry \cite{AST01a} is
characterized by the so-called type A monomial space \cite{GT05}
preserved by $\tH_{\cN}^{-}$:
\begin{align}
\tcV_{\cN}=\braket{1,z,\dots,z^{\cN-1}}.
\label{eq:Ams}
\end{align}
Applying the algorithm with this type A monomial space, we can
construct the most general form of type A $\cN$-fold
supersymmetric PDM quantum systems. Fortunately, we can omit
the process of constructing gauged Hamiltonians. Noting the fact
that the form of gauged Hamiltonians $\bar{\tH}_{\cN}^{\pm}$ is
completely the same as in the constant-mass case, we immediately
have \cite{Ta03a}
\begin{align}
\bar{\tH}_{\cN}^{\pm}=&\,-A(z)\frac{\rmd^{2}}{\rmd z^{2}}+\biggl[
 \frac{\cN-2}{2}A'(z)\pm Q(z)\biggr]\frac{\rmd}{\rmd z}\notag\\
&\,-\biggl[\frac{(\cN-1)(\cN-2)}{12}A''(z)\pm\frac{\cN-1}{2}
 Q'(z)+R\biggr],
\label{eq:gHamA}
\end{align}
where $R$ is a constant, and $Q(z)$ and $A(z)$ must satisfy
\begin{align}
\frac{\rmd^{3}Q(z)}{\rmd z^{3}}&=0\quad\text{for}\quad\cN\geq 2,
\label{eq:condQ}\\
\frac{\rmd^{5}A(z)}{\rmd z^{5}}&=0\quad\text{for}\quad\cN\geq 3,
\label{eq:condA}
\end{align}
or equivalently,
\begin{alignat}{2}
Q(z)&=b_{2}z^{2}+b_{1}z+b_{0} & \quad\text{for}\quad\cN&\geq 2,\\
A(z)&=a_{4}z^{4}+a_{3}z^{3}+a_{2}z^{2}+a_{1}z+a_{0}
 & \quad\text{for}\quad\cN&\geq 3,
\end{alignat}
where $b_{i}$ and $a_{i}$ are constants. The most general
$\cN$th-order linear differential operator of the form \eqref{eq:tPN-},
with $g(z)$ given by Eq.~\eqref{eq:gofz}, which annihilates the type A
space is obviously
\begin{align}
\tP_{\cN}^{-}=m(q)^{-\frac{\cN}{2}}z'(q)^{\cN}
 \frac{\rmd^{\cN}}{\rmd z^{\cN}}.
\label{eq:AtPN-}
\end{align}
Hence, using Eqs.~\eqref{eq:drvPN}, \eqref{eq:defEq}, and
\eqref{eq:cWN+-}, we obtain the operator $P_{\cN}$ for the type A
$\cN$-fold supercharge
\begin{align}
P_{\cN}=m(q)^{-\frac{\cN}{2}}\prod_{k=0}^{\cN-1}\left(
 \frac{\rmd}{\rmd q}+W(q)-\frac{\cN m'(q)}{4m(q)}
 +\frac{\cN-1-2k}{2}E(q)\right),
\label{eq:APN}
\end{align}
where the products of operators are ordered according to the
following definition:
\begin{align}
\prod_{k=k_{0}}^{k_{1}}A_{k}=A_{k_{1}}A_{k_{1}-1}\dots A_{k_{0}}.
\end{align}
The above $P_{\cN}$ indeed reduces to the ordinary type A $\cN$-fold
supercharge when $m(q)=1$. The pair of type A PDM Hamiltonians
$H_{\cN}^{\pm}$ can be expressed in terms of the functions $E(q)$,
$W(q)$, and $m(q)$ using the following formulas, which follow from
Eqs.~\eqref{eq:cond3} and \eqref{eq:defQz}:
\begin{align}
A'(z)&=z'(q)\left(\frac{E(q)}{m(q)}-\frac{m'(q)}{2m(q)^{2}}\right),\\
A''(z)&=\frac{E'(q)+E(q)^{2}}{m(q)}-\frac{3m'(q)E(q)}{2m(q)^{2}}
 -\frac{m(q)m''(q)-2m'(q)^{2}}{2m(q)^{3}},\\
Q'(z)&=-\frac{W'(q)+E(q)W(q)}{m(q)}+\frac{m'(q)W(q)}{m(q)^{2}}.
\end{align}
From Eqs.~\eqref{eq:sympH}, \eqref{eq:cWN+-}, \eqref{eq:gHamA},
and the above formulas, we finally obtain
\begin{align}
H_{\cN}^{\pm}=&\,-\frac{1}{2m(q)}\frac{\rmd^{2}}{\rmd q^{2}}
 +\frac{m'(q)}{2m(q)^{2}}\frac{\rmd}{\rmd q}+\frac{W(q)^{2}}{2m(q)}
 -\frac{\cN^{\,2}-1}{24m(q)}\bigl(2E'(q)-E(q)^{2}\bigr)\notag\\
&\,+\frac{\cN^{\,2}+2}{24}\frac{m''(q)}{m(q)^{2}}
 -\frac{5\cN^{\,2}+16}{96}\frac{m'(q)^{2}}{m(q)^{3}}
\pm\cN\left(\frac{W'(q)}{2m(q)}-\frac{m'(q)W(q)}{4m(q)^{2}}\right)-R.
\label{eq:AH+-}
\end{align}
This is the most general form of type A $\cN$-fold supersymmetric
PDM Hamiltonians and exactly reduces to the ordinary type A when
$m(q)=1$. It should be also noted that the above formula is
consistent with Eq.~\eqref{eq:diffH}. From the form of the type
A $\cN$-fold supercharge \eqref{eq:APN} we have
\begin{align}
w_{\cN-1}(q)=\cN m(q)^{-\frac{\cN}{2}}W(q)-\frac{\cN^{\,2}}{4}
 m(q)^{-\frac{\cN+2}{2}}m'(q).
\end{align}
With this formula, it is straightforward to check for the type A
Hamiltonians that
\begin{align*}
H_{\cN}^{+}-H_{\cN}^{-}&=\frac{\cN W'}{m}-\frac{\cN m'W}{2m^{2}}\\
&=m^{\frac{\cN-2}{2}}w'_{\cN-1}+\frac{\cN-1}{2}
 m^{\frac{\cN-4}{2}}m'w_{\cN-1}+\frac{\cN^{\,2}m''}{4m^{2}}
 -\frac{3\cN^{\,2}m'^{2}}{8m^{3}},
\end{align*}
and thus consistent with Eq.~\eqref{eq:diffH}. Finally, we shall
express the two conditions \eqref{eq:condQ} and \eqref{eq:condA}
for type A $\cN$-fold supersymmetry in terms of $E(q)$, $W(q)$, and
$m(q)$. Using Eqs.~\eqref{eq:cond3}, \eqref{eq:defEq}, and
\eqref{eq:defQz}, we obtain
\begin{alignat}{2}
\left(\frac{\rmd}{\rmd q}-E\right)\frac{\rmd}{\rmd q}\left(
 \frac{\rmd}{\rmd q}+E\right)\frac{W}{m}&=0 & \quad
 \text{for}\quad\cN&\geq 2,\\
\left(\frac{\rmd}{\rmd q}-2E\right)\left(\frac{\rmd}{\rmd q}
 -E\right)\frac{\rmd}{\rmd q}\left(\frac{\rmd}{\rmd q}+E
 \right)\left(\frac{E}{m}-\frac{m'}{2m^{2}}\right)
 &=0 & \quad\text{for}\quad\cN&\geq 3.
\end{alignat}
In particular, in the case of $\cN=1$ the system composed of
$P_{\cN}$ and $H_{\cN}^{\pm}$ reduces to
\begin{align}
P_{1}=&\,\frac{1}{m(q)^{\frac{1}{2}}}\left(\frac{\rmd}{\rmd q}
 +W(q)-\frac{m'(q)}{4m(q)}\right),
\label{eq:fP1}\\
H_{1}^{\pm}=&\,-\frac{1}{2m(q)}\frac{\rmd^{2}}{\rmd q^{2}}
 +\frac{m'(q)}{2m(q)^{2}}\frac{\rmd}{\rmd q}
 +\frac{W(q)^{2}}{2m(q)}+\frac{m''(q)}{8m(q)^{2}}
 -\frac{7m'(q)^{2}}{32m(q)^{3}}\notag\\
&\,\pm\left(\frac{W'(q)}{2m(q)}-\frac{m'(q)W(q)}{4m(q)^{2}}\right)-R,
\label{eq:fH1+-}
\end{align}
with which the super-Hamiltonian $\bH_{1}$ and supercharges
$\bQ_{1}^{\pm}$ defined by Eqs.~\eqref{eq:defsH} and \eqref{eq:defQN}
satisfy ordinary superalgebra:
\begin{align}
\bigl[\bQ_{1}^{\pm},\bH_{1}\bigr]=0,\quad
 \bigl\{\bQ_{1}^{-},\bQ_{1}^{+}\bigr\}=2(\bH_{\cN}+R).
\end{align}
Hence, it exactly reduces to (ordinary) supersymmetric PDM quantum
systems. We note that thanks to the assignment of the functions $W(q)$
and $m(q)$ in Eq.~\eqref{eq:fP1}, which is different from the one
conventionally employed in the literature, the resulting pair of
$H_{1}^{\pm}$ has completely the symmetric form.

From Eqs.~\eqref{eq:bPN+} and \eqref{eq:AtPN-} the operator
$\bar{P}_{\cN}^{+}$ in the type A case reads
\begin{align}
\bar{P}_{\cN}^{+}=(-1)^{\cN}m(q)^{-\frac{\cN}{2}}z'(q)^{\cN}
 \frac{\rmd^{\cN}}{\rmd z^{\cN}},
\end{align}
and thus $\bar{\cV}_{\cN}^{+}=\ker\bar{P}_{\cN}^{+}$ is also the
type A monomial space \eqref{eq:Ams}. Hence, the solvable sectors
of the type A Hamiltonians $H_{\cN}^{\pm}$ are given by
\begin{align}
\cV_{\cN}^{\pm}=\rme^{-\cW_{\cN}^{\pm}}\braket{1,z,\dots,z^{\cN-1}}
\bigr|_{z=z(q)},
\label{eq:solA1}
\end{align}
where $\cW_{\cN}^{\pm}$ are defined in Eq.~\eqref{eq:cWN+-}.

Another important consequence of the fact that the type A systems
with arbitrary PDM are obtained from the same gauged Hamiltonians
\eqref{eq:gHamA} is that we can obtain completely the same results
as in the constant mass case if we follow the procedure of
generating the generalized Bender--Dunne polynomial (GBDP) systems
$\{\pi_{n}^{[\cN]}(E)\}_{n=0}^{\infty}$ \cite{Ta03a,BD96}.
Hence, the anti-commutator of the type A $\cN$-fold supercharges
in the PDM case is also proportional to the $\cN$th critical
GBDP in the type A super-Hamiltonian (for details see
Ref.~\cite{Ta03a} and references cited therein):
\begin{align}
\bigl\{\bQ_{\cN}^{-},\bQ_{\cN}^{+}\bigr\}=2^{\cN}\pi_{\cN}^{[\cN]}
 (\bH_{\cN}).
\end{align}

\section{Classification of the Models}
\label{sec:class}

Owing to the fact that the form of the gauged Hamiltonians does not
depend on the mass function $m(q)$, we can classify type A $\cN$-fold
supersymmetric PDM systems completely the same way as in the
constant mass case \cite{Ta03a}. In the complex classification scheme
based on the $GL(2,\bbC)$ invariance, there are five inequivalent
models according to different canonical forms of $A(z)$ given in
the second column of Table~\ref{tb:canon}.

\tabcolsep=10pt
\begin{table}
\begin{center}
\begin{tabular}{lll}
\hline
Case & Canonical Form & $f(u)$\\
\hline
I   & $1/2$                     & $u$\\
II  & $2z$                      & $u^{2}$\\
III & $2\nu z^{2}$              & $\rme^{2\rnu u}$\\
IV  & $2\nu(z^{2}-1)$           & $\cosh 2\rnu u$\\
V   & $2z^{3}-g_{2}z/2-g_{3}/2$ & $\wp(u)$\\
\hline
\end{tabular}
\caption{Canonical forms of $A(z)$ and the functions $f(u)$ which
characterize the change of variable. The parameters $\nu, g_{2},
g_{3}\in\bbC$ satisfy $\nu\neq 0$ and $g_{2}^{3}-27g_{3}^{2}\neq 0$.}
\label{tb:canon}
\end{center}
\end{table}

A new feature due to the position dependence of mass emerges
through the change of variable. From Eq.~\eqref{eq:cond3} the
change of variable is determined by
\begin{align}
\pm u(q)\equiv\pm\int\rmd q\,\sqrt{m(q)}=\int\frac{\rmd z}{
 \sqrt{2A(z)}}.
\end{align}
Expressing the variable $z$ in terms of $u$ from this formula,
$z=f(u)$, we obtain the change of variable as
\begin{align}
z=z(q)=f(u)\bigr|_{u=u(q)}.
\label{eq:deffu}
\end{align}
The function $f(u)$ in each of the five cases is given in the third
column of Table~\ref{tb:canon}. When $m(q)=1$, it is evident that
$u(q)=q$ (up to an additive constant), and thus reproduces all
the type A models with constant mass in Ref.~\cite{Ta03a}. The form of
the change of variable \eqref{eq:deffu} indicates that it is more
convenient to express the type A PDM Hamiltonians \eqref{eq:AH+-} in
terms of $f(u)$. To this end, we must first express derivatives of
$z$ with respect to $q$ in terms of $f(u)$ and $m(q)$. For instance,
the first derivative of $z(q)$ reads
\begin{align}
z'(q)=f'(u)u'(q)=m(q)^{\frac{1}{2}}f'(u).
\label{eq:1dz}
\end{align}
Similarly, we can derive
\begin{align}
z''(q)&=m(q)f''(u)+\frac{m'(q)}{2m(q)^{\frac{1}{2}}}f'(u),
\label{eq:2dz}\\
z'''(q)&=m(q)^{\frac{3}{2}}f'''(u)+\frac{3}{2}m'(q)f''(u)
 +\biggl(\frac{m''(q)}{2m(q)^{\frac{1}{2}}}-\frac{m'(q)^{2}}{
 4m(q)^{\frac{3}{2}}}\biggr)f'(u).
\label{eq:3dz}
\end{align}
Using Eqs.~\eqref{eq:defEq} and \eqref{eq:1dz}--\eqref{eq:3dz},
we obtain,
\begin{align}
2E'(q)-E(q)^{2}&=\frac{2z'''(q)}{z'(q)}-\frac{3z''(q)^{2}}{
 z'(q)^{2}}\notag\\
&=m(q)\left(\frac{2f'''(u)}{f'(u)}-\frac{3f''(u)^{2}}{f'(u)^{2}}
 \right)+\frac{m''(q)}{m(q)}-\frac{5m'(q)^{2}}{4m(q)^{2}}.
\label{eq:Efm}
\end{align}
The function $W(q)$ and its derivative are also expressed in terms
of $f(u)$ and $m(q)$ with the aid of Eqs.~\eqref{eq:defQz},
\eqref{eq:1dz} and \eqref{eq:2dz} as
\begin{align}
W(q)&=-\frac{m(q)^{\frac{1}{2}}}{f'(u)}Q(z)\biggl|_{z=f(u)},\\
W'(q)&=\left[\biggl(\frac{m(q)f''(u)}{f'(u)^{2}}-\frac{m'(q)}{
 2m(q)^{\frac{1}{2}}f'(u)}\biggr)Q(z)-m(q)Q'(z)\right]_{z=f(u)}.
\label{eq:W'fm}
\end{align}
Substituting Eqs.~\eqref{eq:Efm}--\eqref{eq:W'fm} into
Eq.~\eqref{eq:AH+-}, we finally have the expression of the type A
PDM Hamiltonians in terms of $f(u)$ and $m(q)$ as follows:
\begin{align}
H_{\cN}^{\pm}=&\,-\frac{1}{2m(q)}\frac{\rmd^{2}}{\rmd q^{2}}
 +\frac{m'(q)}{m(q)^{2}}\frac{\rmd}{\rmd q}+\Biggl[
 \frac{Q(z)^{2}}{2f'(u)^{2}}-\frac{\cN^{\,2}-1}{24}\biggl(
 \frac{2f'''(u)}{f'(u)}-\frac{3f''(u)^{2}}{f'(u)^{2}}\biggr)\notag\\
&\,+\frac{m''(q)}{8m(q)^{2}}-\frac{7m'(q)^{2}}{32m(q)^{3}}
 \pm\frac{\cN}{2}\biggl(\frac{f''(u)}{f'(u)^{2}}Q(z)-Q'(z)\biggr)
 -R\Biggr]_{z=f(u)}.
\end{align}
Similarly, the gauge potentials $\cW_{\cN}^{\pm}$ are expressed as
\begin{align}
\cW_{\cN}^{\pm}=-\frac{1}{4}\ln |m(q)|+\frac{\cN-1}{2}\ln |f'(u)|
 \pm\int\rmd u\,\frac{Q(f(u))}{f'(u)}.
\end{align}
Hence, the solvable sectors of the type A Hamiltonians
\eqref{eq:solA1} reads
\begin{align}
\cV_{\cN}^{\pm}=m(q)^{\frac{1}{4}}f'(u)^{-\frac{\cN-1}{2}}
 \exp\left(\mp\int\rmd u\frac{Q(f(u))}{f'(u)}\right)
 \braket{1,f(u),\dots,f(u)^{\cN-1}}.
\end{align}
Observing the above formulas and noting the fact that $u(q)\to q$
(up to an additive constant) as $m(q)\to 1$, we can find a simple
procedure to obtain a type A PDM quantum system from a type A
constant-mass model. That is, if we denote the pair of potentials,
gauge potentials, and solvable sectors of a type A constant-mass
model as $V_{\cN}^{(0)\pm}(q)$, $\cW_{\cN}^{(0)\pm}(q)$, and
$\cV_{\cN}^{(0)\pm}[q]$, respectively, those of the corresponding
type A PDM model are given by
\begin{subequations}
\label{eqs:pdmcm}
\begin{align}
U_{\cN}^{\pm}(q)&=V_{\cN}^{(0)\pm}(u(q))+\frac{m''(q)}{8m(q)^{2}}
 -\frac{7m'(q)^{2}}{32m(q)^{3}},\\
\cW_{\cN}^{\pm}(q)&=-\frac{1}{4}\ln |m(q)|+\cW_{\cN}^{(0)\pm}(u(q)),
 \quad\cV_{\cN}^{\pm}[q]=m(q)^{\frac{1}{4}}\cV_{\cN}^{(0)\pm}[u(q)].
\end{align}
\end{subequations}
This result is consistent with the one obtained by the point canonical
transformation, see e.g., Eqs.~(2.7) and (2.8) in Ref.~\cite{Al02},
and Eqs.~(10), (13) and (14) in Ref.~\cite{GOGU02}.
Advantages of the framework of $\cN$-fold supersymmetry are that
it can automatically determine the general functional form of
the potentials for which we can obtain (a number of) analytic
solutions and that, in a good situation like the present type A case,
we can completely classify all the possible (quasi-)solvable models
which have a specific type of solutions. In addition, it enables us
to obtain simultaneously a pair of almost isospectral Hamiltonians.

In what follows, we show the explicit form of effective potentials
and solvable sectors in each case classified as in Table~\ref{tb:canon}.
We note that type A models become not only quasi-solvable but also
solvable when the parameters in the potentials satisfy
\begin{align}
a_{4}=a_{3}=b_{2}=0.
\end{align}
Hence, the models in Cases I--IV are solvable when $b_{2}=0$, while
the model in Case V is always only quasi-solvable irrespective of
the values of the parameters $b_{i}$.

\subsection{Case I: $A(z)=1/2$, $f(u)=u$}
\label{ssec:case1}

\noindent
\emph{Effective potentials:}
\begin{align}
U_{\cN}^{\pm}(q)=\frac{1}{2}\left(b_{2}u(q)^{2}+b_{1}u(q)+b_{0}
 \right)^{2}\mp\cN b_{2}u(q)+\frac{m''(q)}{8m(q)^{2}}
 -\frac{7m'(q)^{2}}{32m(q)^{3}}\mp\frac{\cN b_{1}}{2}-R.
\end{align}
\noindent
\emph{Solvable sectors:}
\begin{align}
\cV_{\cN}^{\pm}=m(q)^{\frac{1}{4}}\exp\left(\mp\frac{b_{2}}{3}
 u(q)^{3}\mp\frac{b_{1}}{2}u(q)^{2}\mp b_{0}u(q)\right)
\braket{1,u(q),\dots,u(q)^{\cN-1}}.
\end{align}

\subsection{Case II: $A(z)=2z$, $f(u)=u^{2}$}
\label{ssec:case2}

\noindent
\emph{Effective potentials:}
\begin{align}
U_{\cN}^{\pm}(q)=&\,\frac{b_{2}^{2}}{8}u(q)^{6}+\frac{b_{2}b_{1}}{4}
 u(q)^{4}+\frac{1}{8}(b_{1}^{2}+2b_{0}b_{2}\mp 6\cN b_{2})u(q)^{2}
 \notag\\
&\,+\frac{(\cN-1\pm b_{0})(\cN+1\pm b_{0})}{8u(q)^{2}}
 +\frac{m''(q)}{8m(q)^{2}}-\frac{7m'(q)^{2}}{32m(q)^{3}}
 \mp\frac{\cN b_{1}}{4}+\frac{b_{0}b_{1}}{4}-R.
\end{align}
\noindent
\emph{Solvable sectors:}
\begin{align}
\cV_{\cN}^{\pm}=m(q)^{\frac{1}{4}}u(q)^{-\frac{\cN-1\pm b_{0}}{2}}
 \exp\left(\mp\frac{b_{2}}{8}u(q)^{4}\mp\frac{b_{1}}{4}u(q)^{2}\right)
\braket{1,u(q)^{2},\dots,u(q)^{2(\cN-1)}}.
\end{align}

\subsection{Case III: $A(z)=2\nu z^{2}$, $f(u)=\rme^{2\rnu u}$}
\label{ssec:case3}

\noindent
\emph{Effective potentials:}
\begin{align}
U_{\cN}^{\pm}(q)=&\,\frac{b_{2}^{2}}{8\nu}\rme^{4\rnu u(q)}
 +\frac{b_{2}}{4\nu}(b_{1}\mp 2\cN\nu)\rme^{2\rnu u(q)}
 +\frac{b_{0}}{4\nu}(b_{1}\pm 2\cN\nu)\rme^{-2\rnu u(q)}
 +\frac{b_{0}^{2}}{8\nu}\rme^{-4\rnu u(q)}\notag\\
&\,+\frac{m''(q)}{8m(q)^{2}}-\frac{7m'(q)^{2}}{32m(q)^{3}}
 +\frac{b_{1}^{2}+2b_{2}b_{0}}{8\nu}+\frac{\cN^{\,2}-1}{6}\nu-R.
\end{align}
\noindent
\emph{Solvable sectors:}
\begin{align}
\cV_{\cN}^{\pm}=&\,m(q)^{\frac{1}{4}}\exp\left(\mp\frac{b_{2}}{4\nu}
 \rme^{2\rnu u(q)}\pm\frac{b_{0}}{4\nu}\rme^{-2\rnu u(q)}
 -\frac{2(\cN-1)\nu\pm b_{1}}{2\rnu}u(q)\right)\notag\\
&\,\times\braket{1,\rme^{2\rnu u(q)},\dots,\rme^{2(\cN-1)\rnu u(q)}}.
\end{align}

\subsection{Case IV: $A(z)=2\nu(z^{2}-1)$, $f(u)=\cosh 2\rnu u$}
\label{ssec:case4}

\noindent
\emph{Effective potentials:}
\begin{align}
U_{\cN}^{\pm}(q)=&\,\frac{b_{2}^{2}}{8\nu}\sinh^{2}2\rnu u(q)
 +\frac{b_{2}(b_{1}\mp 2\cN\nu)}{2\nu}\sinh^{2}\rnu u(q)
 +\frac{(b_{2}+b_{0})(b_{1}\pm 2\cN\nu)}{8\nu\sinh^{2}\rnu u(q)}\notag\\
&\,+\frac{(b_{2}+b_{0}-b_{1}\mp 2(\cN-1)\nu)(b_{2}+b_{0}-b_{1}
 \mp 2(\cN+1)\nu)}{8\nu\sinh^{2}2\rnu u(q)}\notag\\
&\,+\frac{m''(q)}{8m(q)^{2}}-\frac{7m'(q)^{2}}{32m(q)^{3}}
 \mp\frac{\cN b_{2}}{2}+\frac{2b_{2}(b_{2}+b_{0}+b_{1})+b_{1}^{2}}{
 8\nu}+\frac{\cN^{\,2}-1}{6}\nu-R.
\end{align}
\noindent
\emph{Solvable sectors:}
\begin{align}
\cV_{\cN}^{\pm}=&\,m(q)^{\frac{1}{4}}(\sinh 2\rnu u(q))^{-\frac{\cN-1}{2}
 \mp\frac{b_{1}}{4\nu}}(\tanh \rnu u(q))^{\mp\frac{b_{2}+b_{0}}{4\nu}}
 \exp\left(\mp\frac{b_{2}}{4\nu}\cosh 2\rnu u(q)\right)\notag\\
&\,\times\braket{1,\cosh 2\rnu u(q),\dots,(\cosh 2\rnu u(q))^{\cN-1}}.
\end{align}

\subsection{Case V: $A(z)=2z^{3}-g_{2}z/2-g_{3}/2$, $f(u)=\wp(u)$}
\label{ssec:case5}

\noindent
\emph{Effective potentials:}
\begin{align}
U_{\cN}^{\pm}(q)=&\,\sum_{l=1}^{3}\frac{\eta_{l}^{\pm}}{8H_{l}^{2}
 [\wp(u(q))-e_{l}]}+\frac{(\cN-1\mp b_{2})(\cN+1\mp b_{2})}{8}
 \wp(u(q))\notag\\
&\,+\frac{m''(q)}{8m(q)^{2}}-\frac{7m'(q)^{2}}{32m(q)^{3}}
 \pm\frac{\cN b_{1}}{4}+\frac{b_{2}b_{1}}{4}-R,
\end{align}
where $e_{l}=\wp(w_{l})$ ($l=1,2,3$) are the values of the
Weierstrass function at the half of the fundamental periods $2w_{l}$
and $H_{l}^{2}=3e_{l}^{2}-g_{2}/4$. The coupling constants
$\eta_{l}^{\pm}$ are given by
\begin{align}
\eta_{l}^{\pm}=&\,-b_{2}e_{l}(b_{2}e_{l}-2b_{1})(2H_{l}^{2}-5e_{l}^{2})
 +(b_{1}^{2}+2b_{2}b_{0})e_{l}^{2}-2b_{1}b_{0}e_{l}+b_{0}^{2}\notag\\
&\,+(\cN^{\,2}-1)(H_{l}^{4}-18e_{l}^{2}H_{l}^{2}+36e_{l}^{4})
 \mp 2\cN\bigl[(b_{2}e_{l}-b_{1})(5H_{l}^{2}-12e_{l}^{2})e_{l}
 -b_{0}H_{l}^{2}\bigr].
\end{align}
\noindent
\emph{Solvable sectors:}
\begin{align}
\cV_{\cN}^{\pm}=&\,m(q)^{\frac{1}{4}}\prod_{l=1}^{3}\bigl|\wp(u(q))-
 e_{l}\bigr|^{-\frac{\cN-1}{4}\mp\frac{b_{2}e_{l}^{2}-b_{1}e_{l}
 +b_{0}}{4H_{l}^{2}}}
\braket{1,\wp(u(q)),\dots,\wp(u(q))^{\cN-1}}.
\end{align}

\section{Discussion and Summary}
\label{sec:discus}

In this paper, we have generalized $\cN$-fold supersymmetry in
ordinary quantum systems to those with position-dependent mass.
The significant properties such as the equivalence to weak
quasi-solvability also hold in the PDM case. We have developed
the general procedure to construct an $\cN$-fold supersymmetric
PDM system and applied it to obtain the general form of type A
$\cN$-fold supersymmetry in PDM quantum systems. it turns out
that the framework of $\cN$-fold supersymmetry is quite powerful
also in searching (quasi-)solvable PDM Hamiltonians as well
as ordinary ones. In fact, many of the so far constructed
(quasi-)solvable PDM Hamiltonians in the literature are realized
as special cases of type A $\cN$-fold supersymmetry. In addition,
we can simultaneously obtain a pair of almost isospectral PDM
Hamiltonians in the framework of $\cN$-fold supersymmetry.

There are a lot of future issues worthy investigating in this
research direction. For example, it would be straightforward to
construct other types of $\cN$-fold supersymmetry, namely,
type B \cite{GT04} and type C \cite{GT05} in PDM quantum systems.
They are characterized by respectively the type B monomial space
\begin{align}
\tcV_{\cN}=\braket{1,z,\dots,z^{\cN-2},z^{\cN}},
\end{align}
and by the type C monomial space
\begin{align}
\tcV_{\cN}=\braket{1,z,\dots,z^{\cN_{1}-1},z^{\lambda},
 z^{\lambda+1},\dots,z^{\lambda+\cN_{2}-1}},
\end{align}
with $\lambda\in\bbR\setminus\{-\cN_{2},-\cN_{2}+1,\dots,\cN_{1}\}$.
Applying the algorithm in Section~\ref{sec:algo} with above spaces,
we will obtain the general form of type B and type C $\cN$-fold
supersymmetric PDM quantum systems. We expect the same relations as
Eqs.~\eqref{eqs:pdmcm} since they are universal in view of the
point canonical transformation.

Investigation of dynamical properties are interesting and important.
In contrast to the ordinary constant-mass case, not only the form of
potential but also the position dependence of mass affect various
aspects of PDM systems. In particular, dynamical $\cN$-fold
supersymmetry breaking can take place through the nonperturbative
effect due to quantum tunneling \cite{AKOSW99,ST02}. Hence, it is
quite interesting if we can experimentally observe such a phenomenon
in realistic systems such as semiconductors, quantum dots, and so on.

Construction of (quasi-)solvable quantum many-body systems
with position-dependent mass is a challenging problem. Recently,
we have developed systematic and powerful methods for constructing
quasi-solvable differential operators of arbitrary number of variables
and applied them to construct quasi-solvable many-body Hamiltonians
with constant mass \cite{Ta04,Ta05a,Ta05d}. We expect that the methods
would also work well in the PDM case. Up to now, we have found four
different types of quasi-solvable differential operators of arbitrary
number of variables, namely, type A \cite{Ta04}, type C \cite{Ta05a},
type A$'$, and type C$'$ \cite{Ta05d}. In the constant-mass case, most
of the obtained Hamiltonians are Calogero--Sutherland and Inozemtsev
models associated with classical root systems. Hence, we will obtain
a set of mass-deformed quantum systems of these types which preserve
(quasi-)solvability. In this respect, we anticipate a many-body
generalization of the relations \eqref{eqs:pdmcm}, which relates
a (quasi-)solvable potential in PDM systems with that in constant-mass
systems.

\begin{acknowledgments}
 This work was partially supported by the National Science Council
 of the Republic of China under grant No. NSC-93-2112-M-032-009.
\end{acknowledgments}


\bibliography{refsels}

\begin{thebibliography}{10}
\expandafter\ifx\csname url\endcsname\relax
  \def\url#1{{\tt #1}}\fi
\expandafter\ifx\csname urlprefix\endcsname\relax\def\urlprefix{URL }\fi
\providecommand{\eprint}[2][]{\url{#2}}

\bibitem{DCH98}
L.~Dekar, L.~Chetouani, and T.~F. Hammann, J. Math. Phys. 39 (1998) 2551.

\bibitem{DCH99}
L.~Dekar, L.~Chetouani, and T.~F. Hammann, Phys. Rev. A 59 (1999) 107.

\bibitem{MI99}
V.~Milanovi{\'c} and Z.~Ikoni{\'c}, J. Phys. A: Math. Gen. 32 (1999) 7001.

\bibitem{PRCGP99}
A.~R. Plastino, A.~Rigo, M.~Casas, F.~Garcias, and A.~Plastino, Phys. Rev. A 60
  (1999) 4318.

\bibitem{DA00}
A.~de~Souza~Dutra and C.~A.~S. Almeida, Phys. Lett. A 275 (2000) 25.
\newblock \eprint{quant-ph/0306065}.

\bibitem{Al02}
A.~D. Alhaidari, Phys. Rev. A 66 (2002) 042116.
\newblock \eprint{quant-ph/0207061}.

\bibitem{GGTO02}
B.~G{\"o}n{\"o}l, B.~G{\"o}n{\"o}l, D.~Tutcu, and O.~{\"O}zer, Mod. Phys. Lett.
  A 17 (2002) 2057.
\newblock \eprint{quant-ph/0211112}.

\bibitem{GOGU02}
B.~G{\"o}n{\"o}l, O.~{\"O}zer, B.~G{\"o}n{\"o}l, and F.~{\"U}zg{\"u}n, Mod.
  Phys. Lett. A 17 (2002) 2453.
\newblock \eprint{quant-ph/0211113}.

\bibitem{KKK02}
R.~Ko{\c c}, M.~Koca, and E.~K{\"o}rc{\"u}k, J. Phys. A: Math. Gen. 35 (2002)
  L527.
\newblock \eprint{quant-ph/0410108}.

\bibitem{RR02}
B.~Roy and P.~Roy, J. Phys. A: Math. Gen. 35 (2002) 3961.

\bibitem{DHA03}
A.~de~Souza~Dutra, M.~Hott, and C.~A.~S. Almeida, Europhys. Lett. 62 (2003) 8.
\newblock \eprint{hep-th/0306078}.

\bibitem{KK03}
R.~Ko{\c c} and M.~Koca, J. Phys. A: Math. Gen. 36 (2003) 8105.
\newblock \eprint{quant-ph/0410127}.

\bibitem{KT03}
R.~Ko{\c c} and H.~T{\"u}t{\"u}nc{\"u}ler, Ann. Phys. (Leipzig) 12 (2003) 684.
\newblock \eprint{quant-ph/0410088}.

\bibitem{SL03}
K.~A. Samani and F.~Loran.
\newblock Shape invariant potentials for effective mass {S}chr{\"o}dinger
  equation.
\newblock Preprint, \eprint{quant-ph/0302191}.

\bibitem{YDS04}
J.~Yu, S.-H. Dong, and G.-H. Sun, Phys. Lett. A 322 (2004) 290.

\bibitem{YD04}
J.~Yu and S.-H. Dong, Phys. Lett. A 325 (2004) 194.

\bibitem{Ch04}
G.~Chen, Phys. Lett. A 329 (2004) 22.

\bibitem{CC04}
G.~Chen and Z.~Chen, Phys. Lett. A 331 (2004) 312.

\bibitem{QT04}
C.~Quesne and V.~M. Tkachuk, J. Phys. A: Math. Gen. 37 (2004) 4267.
\newblock \eprint{math-ph/0403047}.

\bibitem{BGQR04a}
B.~Bagchi, P.~Gorain, C.~Quesne, and R.~Roychoudhury, Mod. Phys. Lett. A 19
  (2004) 2765.
\newblock \eprint{quant-ph/0405193}.

\bibitem{GK04}
B.~G{\"o}n{\"o}l and M.~Ko{\c c}ak, Chin. Phys. Lett. 22 (2005) 2742.
\newblock \eprint{quant-ph/0412161}.

\bibitem{BBQT05}
B.~Bagchi, A.~Banerjee, C.~Quesne, and V.~M. Tkachuk, J. Phys. A: Math. Gen. 38
  (2005) 2929.
\newblock \eprint{quant-ph/0412016}.

\bibitem{DL05}
S.-H. Dong and M.~Lozada-Cassou, Phys. Lett. A 337 (2005) 313.

\bibitem{RR05}
B.~Roy and P.~Roy, Phys. Lett. A 340 (2005) 70.

\bibitem{BGQR05}
B.~Bagchi, P.~Gorain, C.~Quesne, and R.~Roychoudhury, Europhys. Lett. 72 (2005)
  155.
\newblock \eprint{quant-ph/0505171}.

\bibitem{Qu05}
C.~Quesne.
\newblock First-order intertwining operators and position-dependent mass
  {S}chr{\"o}dinger equations in d dimensions.
\newblock Preprint, \eprint{quant-ph/0508216}.

\bibitem{vR83}
O.~von Roos, Phys. Rev. B 27 (1983) 7547.

\bibitem{AIS93}
A.~A. Andrianov, M.~V. Ioffe, and V.~P. Spiridonov, Phys. Lett. A 174 (1993)
  273.
\newblock \eprint{hep-th/9303005}.

\bibitem{AST01b}
H.~Aoyama, M.~Sato, and T.~Tanaka, Nucl. Phys. B 619 (2001) 105.
\newblock \eprint{quant-ph/0106037}.

\bibitem{GT05}
A.~Gonz{\'a}lez-L{\'o}pez and T.~Tanaka, J. Phys. A: Math. Gen. 38 (2005) 5133.
\newblock \eprint{hep-th/0405079}.

\bibitem{TU87}
A.~V. Turbiner and A.~G. Ushveridze, Phys. Lett. A 126 (1987) 181.

\bibitem{Us94}
A.~G. Ushveridze, {Q}uasi-exactly {S}olvable {M}odels in {Q}uantum {M}echanics
  (IOP Publishing, Bristol, 1994).

\bibitem{AST01a}
H.~Aoyama, M.~Sato, and T.~Tanaka, Phys. Lett. B 503 (2001) 423.
\newblock \eprint{quant-ph/0012065}.

\bibitem{Ta03a}
T.~Tanaka, Nucl. Phys. B 662 (2003) 413.
\newblock \eprint{hep-th/0212276}.

\bibitem{Tu94}
A.~V. Turbiner, Contemp. Math. 160 (1994) 263.

\bibitem{AS03}
A.~A. Andrianov and A.~V. Sokolov, Nucl. Phys. B 660 (2003) 25.
\newblock \eprint{hep-th/0301062}.

\bibitem{BD96}
C.~M. Bender and G.~V. Dunne, J. Math. Phys. 37 (1996) 6.
\newblock \eprint{hep-th/9511138}.

\bibitem{GT04}
A.~Gonz{\'a}lez-L{\'o}pez and T.~Tanaka, Phys. Lett. B 586 (2004) 117.
\newblock \eprint{hep-th/0307094}.

\bibitem{AKOSW99}
H.~Aoyama, H.~Kikuchi, I.~Okouchi, M.~Sato, and S.~Wada, Nucl. Phys. B 553
  (1999) 644.
\newblock \eprint{hep-th/9808034}.

\bibitem{ST02}
M.~Sato and T.~Tanaka, J. Math. Phys. 43 (2002) 3484.
\newblock \eprint{hep-th/0109179}.

\bibitem{Ta04}
T.~Tanaka, Ann. Phys. 309 (2004) 239.
\newblock Erratum-ibid. 320 (2005) 257, \eprint{hep-th/0306174}.

\bibitem{Ta05a}
T.~Tanaka, Ann. Phys. 316 (2005) 187.
\newblock \eprint{hep-th/0407275}.

\bibitem{Ta05d}
T.~Tanaka, Ann. Phys. 320 (2005) 199.
\newblock \eprint{hep-th/0502019}.

\end{thebibliography}
\bibliographystyle{npb}



\end{document}